\documentstyle[preprint,aps,epsfig]{revtex}
\begin{document}
\draft
\title{The Vector Current Correlator $<0\vert T(V^3_\mu V^8_\nu )\vert 0>$ \\
To Two Loops in Chiral Perturbation Theory \\
\begin{flushright}
ADP-95-27/T181 \\
hep-ph/9504404
\end{flushright}}
\author{Kim Maltman\cite{byline}}
\address{Department of Mathematics and Statistics, York University,
4700 Keele St., \\ North York, Ontario, Canada M3J 1P3}
\date{\today}
\maketitle
\begin{abstract}
The isospin-breaking correlator of the product of flavor octet vector currents,
$\Pi^{38}_{\mu\nu}(q^2)$
$=i\int d^4x\ \exp (iq.x)\ <0\vert T(V^3_\mu (x)V^8_\nu (0))\vert 0>$,
is computed to
next-to-next-to-leading (two-loop) order in Chiral Perturbation Theory.
Large corrections to both the magnitude and $q^2$-dependence of the
one-loop result are found, and the reasons for the slow convergence
of the chiral series for the correlator given.  The two-loop
expression involves a single ${\cal O}(q^6)$ counterterm, present
also in the two-loop expressions for $\Pi^{33}_{\mu\nu}(q^2)$ and
$\Pi^{88}_{\mu\nu}(q^2)$, which counterterm contributes a constant to
the scalar correlator $\Pi^{38}(q^2)$, defined by
$\Pi^{38}_{\mu\nu}(q^2)\equiv (q_\mu q_\nu -q^2g_{\mu\nu})\Pi^{38}(q^2)$.
The feasibility of extracting the value
of this counterterm from other sources is discussed.  Analysis of
the slope of the correlator with respect to $q^2$ using QCD
sum rules is shown to suggest that, even to two-loop order, the chiral
series for the correlator may not yet be well-converged.
\end{abstract}

\section{Introduction}\label{sec:1}

In the last decade, following the appearance of the classic papers
of Gasser and Leutwyler\onlinecite{ref1,ref2,ref3}~, numerous
treatments of low-energy hadronic properties employing the methods
of Chiral Perturbation Theory (ChPT) have appeared (for an excellent
recent review, see Ref.\onlinecite{ref4}).  In the bulk of these
treatments, the chiral expansion has been carried out to next-to-leading
(one-loop) order (i.e. ${\cal O}(q^4)$ in the usual chiral counting).
Expressions for hadronic observables, to this order, incorporate
the constraints of current algebra and, in addition, provide the
leading corrections to these constraints in a transparent and
unambiguous manner.  For many processes (see again Ref.\onlinecite{ref4})
corrections to leading order results are $\simeq 20-30\%$, and
truncating the full chiral series to this order, in consequence, appears
well-justified.  This is, however, not universally the case.
For example, the one-loop amplitude for
$\gamma\gamma\rightarrow\pi^0\pi^0$
\onlinecite{ref5,ref6}, which vanishes at leading order,
differs significantly from experiment even near threshold.  The same
is true of the spectral function of the vector current correlator
$\Pi^{33}_{\mu\nu}(q^2)$, where
\begin{equation}
\Pi^{ab}_{\mu\nu}(q^2)
=i\int d^4x\ \exp (iq.x)\ <0\vert T(V^a_\mu (x)V^b_\nu (0))\vert 0>\equiv
(q_\mu q_\nu -q^2g_{\mu\nu})\Pi^{ab}(q^2)\eqnum{1.1}\label{oneone}
\end{equation}
with $V^a_\mu$ the standard flavor octet vector current,
$V^a_\mu = \bar{q}{\lambda^a\over 2}\gamma_\mu q$, again even
rather near $\pi\pi$ threshold\onlinecite{refGK}~.  Even more
dramatic is the case of the process
$\eta\rightarrow\pi^0\gamma\gamma$,
for which the predicted one-loop branching ratio\onlinecite{ref8,ref9}
is a factor of $\simeq 170$ smaller than the Particle Data Group
\onlinecite{ref10} value.  In the first two cases, the discrepancies between
the one-loop results and experiment are a result of the fact that the
leading order contributions vanish.  Corrections at ${\cal O}(q^6)$
are not unexpectedly large, and recent calculations to two-loop order,
by Bellucci {\it et al.}\onlinecite{refBGS}
for $\gamma\gamma\rightarrow\pi^0\pi^0$,
and by Golowich and Kambor\onlinecite{refGK} for the spectral
functions of $\Pi^{33}_{\mu\nu}$ and $\Pi^{88}_{\mu\nu}$, demonstrate
that inclusion of the ${\cal O}(q^6)$ corrections to the
${\cal O}(q^4)$ one-loop results brings the theoretical predictions
nicely into accord with experiment for $q^2$ less than $\sim (8-9)\,  m_\pi^2$.
The importance of two-loop contributions, even rather near threshold,
has also been demonstrated for the photon vacuum polarization function
in Ref.~\onlinecite{refholdom}.
The situation for $\eta\rightarrow \pi^0\gamma\gamma$ (which
most closely resembles the case at hand)
will be discussed in more detail below.  Other examples of the necessity
of including ${\cal O}(q^6)$ contributions, in the odd intrinsic
parity sector of ${\cal L}_{eff}$, are also known, specifically
$\pi^0\rightarrow\gamma\gamma^*$, $\eta\rightarrow\gamma\gamma^*$
\onlinecite{bijnens90,bijnens91} and $\gamma\pi^+\rightarrow
\pi^+\pi^0$,
$\eta\rightarrow\gamma\pi^+\pi^-$\onlinecite{bijnens90,bijnens91,bijnens93}.

In the present paper we will study the convergence of
the isospin-breaking vector current
correlator, $\Pi^{38}_{\mu\nu}$,
to two-loop order.  We will show that, as for the
chiral series of the amplitude for
$\eta\rightarrow \pi^0\gamma\gamma$, that of the
correlator, $\Pi^{38}_{\mu\nu}$,
is quite poorly converged to one-loop order, and discuss
the physical reasons for this similarity.  We will also discuss
evidence that, even to two-loop order, the latter series
is not yet well-converged.  It should be stressed that the
correlator in question is of interest not only as an example
of a quantity for which the chiral series is slowly converging,
but is also of relevance to ongoing debates concerning the
role of isospin-mixed vector meson exchange in isospin-breaking
and, more particularly, charge-symmetry-breaking, observables
in few-body systems (see Ref.\onlinecite{refKRM} for a discussion
of a number of the contentious issues and list of other
relevant references).  Here the point is that one may choose
the vector currents, rescaled by $g_V/m_V^2$ (where $m_V$, $g_V$
are the corresponding vector meson masses and decay constants,
the latter defined via $<0\vert V^a_\mu \vert V^a(\lambda )>
\equiv {m_V^2\over g_V}\epsilon^\lambda_\mu$ with $a$ the
flavor and $\lambda$ the polarization label of the vector meson)
as interpolating fields for the vector mesons.  The isospin-breaking
correlators $\Pi^{38}_{\mu\nu}$ and $\Pi^{30}_{\mu\nu}$ then
provide information on the $q^2$-dependence of the off-diagonal
elements of the vector meson propagator matrix, for this
choice of interpolating fields.  While the off-shell behavior of
such propagator matrix elements is, in general,
interpolating-field-dependent, one could couple the results
for the propagator to those for the corresponding nucleon-vector meson
vertices, obtained using the same choice of vector meson interpolating
fields, to produce the relevant isospin-breaking contributions to
$NN$ scattering S-matrix elements, such S-matrix elements being
independent of the choice of interpolating fields\onlinecite{ref13}.
Finally, it should also be pointed out that the spectral function
of $\Pi^{38}_{\mu\nu}$ is, at least in principle, measurable
experimentally, though the accuracy required to extract it makes
this a rather moot point, at present.  The possibility of this extraction
rests on the observation that the isovector vector current matrix
elements $<\pi\pi\vert V^k_\mu\vert 0>$ ($k=1,2,3$) receive
isospin-breaking contributions only at second order in $(m_d-m_u)$
\onlinecite{ref14,ref15}, whereas $<\pi\pi\vert V^8_\mu\vert 0>$
is non-zero already at ${\cal O}(m_d-m_u)$.  This means that the
deviation of the ratio of vector spectral functions measured in
$\tau^-\rightarrow\nu_\tau \pi^-\pi^0$ and
$e^+e^-\rightarrow\pi^+\pi^-$
from that predicted by
isospin symmetry is (up to corrections for the
heavy quark pieces of the electromagnetic (EM) current) a direct
measure of the spectral function of $\Pi^{38}_{\mu\nu}$.  Since
these effects will be seen to be of order a few times $10^{-4}$
they are, however, well outside the reach of current experiments,
for which cross-sections below the resonace region in
$e^+e^-\rightarrow\pi^+\pi^-$ are known typically to an
accuracy of only $\simeq 10\%$.

The remainder of the paper is organized as follows.  In Section II
we record the relevant terms of the effective Lagrangian at
orders 2, 4 and 6 in the chiral expansion, and discuss the general,
diagrammatic structure of the one-loop and two-loop results.  In
Section III we describe briefly some details of the calculations and
quote the full one- and two-loop results for the contributions
identified in Section II.  Detailed formulae for the loop integrals
entering these expressions are relegated to the Appendix.  Since
these integrals have been discussed in considerable detail
elsewhere (see, for example Refs.\onlinecite{refGK,refHeathnew}),
the Appendix will be rather brief, and the reader is referred
to the references just cited for further details.
Section IV provides
a discussion of the
results, in particular the physical origin of the slow convergence
of the chiral series for the correlator to one-loop.  In Section V, a rough
estimate of the single ${\cal O}(q^6)$ counterterm appearing in
the ${\cal O}(q^6)$ corrections to the one-loop result
is given, based on a QCD sum rule analysis
of the correlator, and the possibility of independent estimates
of this low-energy constant from other sources discussed.
The issue of the convergence of
the chiral series to two-loop order is also treated.  Finally,
in Section VI, we summarize our conclusions.

\section{The Chiral Lagrangian to ${\cal O}(q^6)$ and the Structure
of the Contributions to the Correlator}\label{sec:2}

The leading terms in the low-energy, chiral expansion of the correlator,
$\Pi^{38}_{\mu\nu}$, may be obtained from the effective chiral
Lagrangian, ${\cal L}_{eff}$, which may be written in the form
\begin{equation}
{\cal L}_{eff}={\cal L}^{(2)}+{\cal L}^{(4)}+{\cal L}^{(6)}+\cdots
\eqnum{2.1}\label{twoone}\end{equation}
where the superscripts denote the chiral order.  The general
form of ${\cal L}^{(2)}$ and ${\cal L}^{(4)}$, in the presence
of external scalar, pseudoscalar, vector and axial vector sources,
is given in Ref.\onlinecite{ref1}~.  Since we are interested only
in the correlator of vector currents we may set the external
pseudoscalar and axial sources to zero and the external scalar
source to $2B_0\, M$, where $M$ is the current quark mass
matrix and $B_0$ the usual parameter, appearing in ${\cal L}^{(2)}$
and related to the value of the quark condensate.  One then has, explicitly,
for ${\cal L}^{(2)}$ and ${\cal L}^{(4)}$\onlinecite{ref1}
\begin{equation}
{\cal L}^{(2)} = {1\over 4}f^2 \text{Tr}(D_\mu U D^\mu
        U^\dagger )
        +{1\over 2}f^2 \text{Tr}[B_0 M(U +U^\dagger )]
\eqnum{2.2}\label{twotwo}\end{equation}
and
\begin{eqnarray}
{\cal L}^{(4)} &=& L_1 \bigl[ \text{Tr}(D_\mu U D^\mu U^\dagger )
        \bigr]^2  \nonumber \\
     && \mbox{} + L_2 \text{Tr}(D_\mu U D_\nu U^\dagger )
        \text{Tr}(D^\mu U D^\nu U^\dagger )
        + L_3 \text{Tr}(D_\mu U^\dagger D^\mu U
        D_\nu U^\dagger D^\nu U)  \nonumber \\
     && \mbox{} + L_4 \text{Tr}(D_\mu U D^\mu U^\dagger )
        \text{Tr}[2B_0 M(U +U^\dagger )]
        + L_5 \text{Tr}\bigl[ 2B_0 (MU +U^\dagger M)
        D_\mu U^\dagger D^\mu U \bigr] \nonumber \\
     && \mbox{} + L_6 \bigl[ \text{Tr}[2B_0
         M(U +U^\dagger )]\bigr]^2
        + L_7 \bigl[ \text{Tr}[2B_0 M(U -U^\dagger )]\bigr]^2
         \nonumber \\
     && \mbox{}+L_8 \text{Tr}[4B_0^2(MU MU +
         MU^\dagger MU^\dagger )]
         -iL_9\text{Tr}[F_{\mu\nu} D^\mu U D^\nu U^\dagger
        +F_{\mu\nu} D^\mu U^\dagger D^\nu U ] \nonumber \\
     && \mbox{} +L_{10}\text{Tr}[U^\dagger F_{\mu\nu}U
         F^{\mu\nu}] +H_1\text{Tr}[F_{\mu\nu}F^{\mu\nu}+
         F_{\mu\nu}F^{\mu\nu}] +H_2\text{Tr}[4B_0^2 M^2]~.
         \eqnum{2.3} \label{twothree}
\end{eqnarray}
In Eqns.~(2.2) and (2.3), $B_0$ is a mass scale
related to the value of the quark condensate in the chiral limit,
$U = \text{exp}(i\vec \lambda \cdot \vec \pi /f) $
(with $\vec\lambda$ the usual $SU(3)$ Gell-Mann matrices and
$\vec\pi$ the octet of pseudoscalar (pseudo-) Goldstone boson fields),
$f$ is a dimensionful
constant, equal to $f_\pi$ in leading order, $M$ is the current
quark mass matrix, and $D_\mu$ is the covariant derivative which,
in the absence of external axial vector sources, takes the form
\begin{equation}
  D_\mu U = \partial_\mu U -i[v_\mu ,U].\eqnum{2.4} \label{twofour}
\end{equation}
The vector field strength tensor, $F_{\mu\nu }$, occuring in Eqn.~(2.3) is
defined by $F_{\mu\nu}=\partial_\mu v_\nu
-\partial_\nu v_\mu - i[v_\mu ,v_\nu ]$,
where  $v_\mu = {\lambda^a\over 2}v^a_\mu$, with $v_\mu^a$ the octet of
external $SU(3)$ vector fields.  For the case at hand we require
only the external sources $v^3_\mu$ and $v^8_\nu$ and hence the last
term in $F_{\mu\nu}$ vanishes.  Note that one would have to supplement
Eqn.~(2.3) with additional terms involving $\text{Tr}(F_{\mu\nu})$ if
one wished to treat the correlator $\Pi^{30}_{\mu\nu}$ but, as
these terms do not enter the calculation of $\Pi^{38}_{\mu\nu}$, we have not
explicitly displayed them in (2.4).  Note also that, in writing the
form (2.3) for ${\cal L}^{(4)}$, additional terms which vanish
as a consequence of the lowest order equation of motion have been
omitted.  In performing calculations to ${\cal O}(q^6)$ one
would, in general, also have to include these terms.  However,
since the effect of their presence can always be absorbed into a
redefinition of the coefficients occuring in ${\cal L}^{(6)}$
\onlinecite{ref4,ref16}, we may drop these terms from the outset.

The general form of ${\cal L}^{(6)}$, in the presence of
external sources, has been determined recently by Fearing
and Scherer\onlinecite{ref17}.  The full expression, however, contains
111 terms of even intrinsic parity and 32 of odd intrinsic parity, and so
will not be recorded here.  In fact, only one combination of
the terms from ${\cal L}^{(6)}$ actually enters the present calculation.
It is easy to see why this is the case.  As pointed out in
Ref.\onlinecite{refGK}, since to ${\cal O}(q^6)$
only terms from ${\cal L}^{(6)}$
zeroth order in the meson fields contribute to vacuum expectations,
and since only four terms of the 143 mentioned
above contain terms zeroth order in the meson fields and second
order in the external vector sources $v^3_\mu$ and $v^8_\nu$,
only these four terms can contribute to the correlators $\Pi^{ab}_{\mu\nu}$,
with $a,b=3,8$.
${\cal L}^{(6)}$ thus reduces, for the purpose of computing such
vacuum correlators to ${\cal O}(q^6)$, to\onlinecite{refGK}
\begin{eqnarray}
{\cal L}^{(6)} &=&{1\over f^2}\Biggl[K_1
\text{Tr}(D_\lambda f_+^{\mu\nu} D^\lambda f_{+\mu\nu})
+K_2\text{Tr}(D^\mu f_+^{\mu\nu} D^\lambda f_{+\lambda\nu})\nonumber \\
&& \mbox{}+K_3\text{Tr}(f_{+\mu\nu}f_+^{\mu\nu}\chi_+)
+K_4\text{Tr}(f_{+\mu\nu}f_+^{\mu\nu})\text{Tr}(\chi_+)\eqnum{2.5}
\label{twofive}\end{eqnarray}
where
\begin{eqnarray}
&& f_+^{\mu\nu} =uv^{\mu\nu}u^\dagger + u^\dagger v^{\mu\nu}u \nonumber \\
&& \chi_+ = 2B_0 u(U^\dagger M+MU)u^\dagger\ .\eqnum{2.6}\label{twosix}
\end{eqnarray}
In Eqns.~(2.5), (2.6), $u=\exp (i \vec\lambda \cdot \vec\pi /2f)$
is the usual square
root of the matrix $U$ defined above, the covariant derivative
of $f_+^{\mu\nu}$ reduces to
\begin{equation}
D_\lambda f_+^{\mu\nu}=\partial_\lambda f_+^{\mu\nu}
-i[v_\lambda ,f_+^{\mu\nu}]\eqnum{2.7}
\label{twoseven}\end{equation}
in the absence of external axial vector sources,
and all other
notation is as defined before.  To zeroth order in the meson
fields, $f_+^{\mu\nu}$ and $\chi_+$ are equal to $2v^{\mu\nu}$ and
$4B_0M$, respectively.  The terms involving $K_1$, $K_2$ and
$K_4$ obviously contain no pieces involving both $v^3_\mu$ and
$v^8_\nu$ and hence do not contribute to the correlator $\Pi^{38}_{\mu\nu}$.
Only the $K_3$ term survives.  To facilitate comparison with
Ref.\onlinecite{refGK} we introduce the rescaled version of
the low-energy constant (LEC) $K_3$, $Q\equiv 4K_3$.

One may now easily characterize the full set of contributions to
the correlator $\Pi^{38}_{\mu\nu}$, to ${\cal O}(q^6)$.
Generically these are of two types, corresponding to the two ways
in which terms involving the product $v^3_\mu v^8_\nu$ can arise
in the expansion of $\exp (i\int d^4x {\cal L}_{eff}[v^a_\mu ])$:
(1) those terms arising from the second order term in the
expansion of the exponential and hence generated by pieces of
${\cal L}_{eff}$ first order in the external vector sources, and
(2) contact terms, arising from the first order term in the
expansion of the exponential, and hence generated by those
pieces of ${\cal L}_{eff}$ second order in the sources.
The resulting contributions to $\Pi^{38}_{\mu\nu}$
are depicted graphically in Figures 1-6.  In these figures
the left-hand current line carries momentum, flavor and
Lorentz indices $q$, $3$ and $\mu$, and the right-hand
current line, similarly, the indices $q$, $8$ and $\nu$.
Open circles enclosing a cross
appearing in Figs.~1-5 denote those vertices generated by
${\cal L}^{(4)}$, and the open box enclosing a cross
in Fig.~6 the vertex
(proportional to $Q$) generated by ${\cal L}^{(6)}$.  All other
vertices are understood to be from ${\cal L}^{(2)}$.
Fig.~1 contains the full set of contributions of ${\cal O}(q^4)$,
Figs.~2-6 those of ${\cal O}(q^6)$.
Figs.~2 and 3 can be interpreted as dressing the internal
propagators of Figs.~1(a) and 1(b).  Additional graphs of
the form 4(a) and 4(b), in which the structures at the
left- and right-hand current vertices have been interchanged
have not been shown explicitly, but are understood to be present.
Since the various vertices appearing in the figures can be read
off from the expressions for ${\cal L}^{(2)}$, ${\cal L}^{(4)}$
and ${\cal L}^{(6)}$, it is a straightforward exercise in
Feynmann diagrammatics to evaluate the correlator.  Results
for the various contributions depicted in the figures are
presented in the next section.

\section{The Correlator $\Pi^{38}_{\mu\nu}$ to
One- and Two-Loop Order}\label{sec:3}

In this section we record the results for the various contributions
to the correlator $\Pi^{38}_{\mu\nu}$,
together with a few salient features of the calculations.
All loop integrals required have been performed using dimensional
regularization and can be expressed in terms of the basic
integrals
\begin{equation}
A(m^2)=\int {d^dk\over (2\pi )^d}{1\over k^2-m^2}\eqnum{3.1}\label{threeone}
\end{equation}
and
\begin{equation}
{\bar B}(m^2,q^2)=-{i\over 16\pi^2}\int^1_0\log (1-q^2x(1-x)/m^2)\ ,
\eqnum{3.2}\label{threetwo}\end{equation}
which are given explicitly in the Appendix.
The auxillary tensors $T_{\mu\nu}(m^2,q^2)$
and $D_{\mu\nu}(m^2,q^2)$, which occur frequently in the calculations
are also described there.
In what follows, the tensor decomposition
\begin{equation}
T_{\mu\nu}(m^2,q^2)=T_1(m^2,q^2)(q_\mu q_\nu -q^2g_{\mu\nu})
+2A(m^2)g_{\mu\nu}\eqnum{3.3}\label{threethree}\\
\end{equation}
and the relation
\begin{equation}
D_{\mu\nu}(m^2,q^2)={1\over 2}T_{\mu\nu}(m^2,q^2) \ ,
\eqnum{3.4}\label{threefour}
\end{equation}
which follows from the expressions given in the Appendix,
have been used to reduce the results to compact forms involving
the integrals $A$, $\bar{B}$ and $T_1$.  The expression
for $T_1$ in terms of $A$ and $\bar{B}$ may also be found
in the Appendix.  In order to streamline
notation we will write $T_1(P)$ for $T_1(m_P^2,q^2)$,
${\bar B}(P)$ for ${\bar B}(m_P^2,q^2)$ and
$A(P)$ for $A(m_P^2)$ in what follows, where $P=K^+$, $K^-$, $\pi$
and $\eta$ and the masses are understood to be those given by
the leading order chiral relations $m_\pi^2=2B_0\hat{m}$,
$m_{K^+}^2=B_0(m_s+m_u)$, $m_{K^0}^2=B_0(m_s+m_d)$ and
$m_\eta^2=2B_0(2m_s+\hat{m})/3$, where $\hat{m}=(m_u+m_d)/2$.

Let us begin with the one-loop, ${\cal O}(q^4)$, result generated by the
diagrams of Fig.~1.  One may easily verify that there are no
contributions to $\Pi^{38}_{\mu\nu}$
of the type 1(c), and that the ${\cal O}(q^4)$ contribution is
generated solely by differences between $K^+$ and $K^0$
loops of types 1(a) and 1(b).  The contributions of type 1(a) are
\begin{equation}
{-i\sqrt{3}\over 4}\left[ T_{\mu\nu}(K^+)-T_{\mu\nu}(K^0)\right]
\eqnum{3.5}\label{threefive}\end{equation}
and those of type 1(b)
\begin{equation}
{-i\sqrt{3}\over 2}\left[  A(K^0) - A(K^+) \right] g_{\mu\nu}\ .
\eqnum{3.6}\label{threesix}\end{equation}
The sum of these contributions yields the full ${\cal O}(q^4)$
result for the correlator,
\begin{equation}
\left[\Pi^{38}_{\mu\nu} (q^2)\right]^{(4)}
={i\sqrt{3}\over 4}
(q_\mu q_\nu -q^2g_{\mu\nu}) \left[
 T_1(K^0) - T_1(K^+) \right]\eqnum{3.7}
\label{threeseven}\end{equation}
which has, of course, the transverse structure required of the vacuum value
of the covariant time ordered product of the conserved vector currents
$V^3_\mu$ and $V^8_\nu$, and can also be seen to be finite and
manifestly independent of the renormalization scale, $\mu$, from the
form of $T_1$ given in the Appendix.

Turning to the contributions of ${\cal O}(q^6)$,
we begin with the insertion graphs of Figs.~2,3.  These are known
in terms of the one-loop contributions to the wavefunction
renormalization constants and mass shifts of the internal
($K^+$ and $K^0$) lines.  The resulting expressions
are considerably simplified if we include also the contributions
of type 4(a) and 5(b) involving the LEC's $L_4$ and $L_5$,
since those contributions exactly cancel the
terms involving explicit factors of $L_4$
and $L_5$ arising from Figs.~2(a) and 3(a).  The resulting
contributions to $\Pi^{38}_{\mu\nu}$ are then
\begin{eqnarray}
(q_\mu q_\nu &&-q^2g_{\mu\nu}) \Biggl(
- {\sqrt{3}\over 24 f^2}
\Biggl[ 3\left(  T_1(K^0) - T_1(K^+) \right)
\left(  A(K^0) + A(K^+) + A(\pi ) + A(\eta ) \right)
\nonumber\\
&&\qquad +\left(  T_1(K^0) + T_1(K^+) \right)
\left( A(K^0) - A(K^+) +2\sqrt{3}\theta_0
\left(  A(\eta ) - A(\pi ) \right)\right)\Biggr]
\nonumber \\
&&\qquad +{i\sqrt{3}\over 2}\left[  (\delta m_{K^+}^2)_{1-loop}
{ \bar{B}(K^+)\over q^2}
- (\delta m_{K^0}^2)_{1-loop} { \bar{B}(K^0)\over q^2}
\right]\Biggr)\nonumber \\
&&+g_{\mu\nu} \Biggl[ - {\sqrt{3}\over 24 f^2} \Biggr]\Biggl[
\left(  A(K^0) - A(K^+) \right)
\left( 4 A(K^+) +4 A(K^0) +3 A(\pi ) +3 A(\eta ) \right)
\nonumber \\
&&\qquad +2\sqrt{3}\theta_0 \left( A(K^0) + A(K^+) \right)
\left( A(\eta ) - A(\pi ) \right)\Biggr]
\eqnum{3.8}\label{threeeight}
\end{eqnarray}
where $\theta_0=\sqrt{3}(m_d-m_u)/4(m_s-\hat{m})$ is
the leading order $\pi$-$\eta$ mixing angle and
$ (\delta m_{K^+}^2)_{1-loop}$, $ (\delta m_{K^0}^2)_{1-loop}$
are the one-loop corrections to the leading order $K^+$ and $K^0$
squared masses, the expressions for which may be found in
Ref.~\onlinecite{ref1}.  The remaining terms of types
4(a) and 5(b), which involve the LEC's $L_{9,10}$, then yield
the contribution
\begin{equation}
(q_\mu q_\nu -q^2g_{\mu\nu})\left[
{-\sqrt{3}\over 24 f^2}\right]\left[
24i\, L_9\, q^2 \left( T_1(K^0) - T_1(K^+) \right)
-48i\, \left( L_9+L_{10}\right)
\left(  A(K^0) - A(K^+) \right)
\right]\ .\eqnum{3.9}\label{threenine}
\end{equation}
The remaining contributions to $\Pi^{38}_{\mu\nu}$
are: (1), from 4(b),
\begin{eqnarray}
(q_\mu q_\nu -q^2g_{\mu\nu}) &&\left[  {\sqrt{3}\over 24 f^2} \right]\Biggl[
\left(  T_1(K^0) - T_1(K^+) \right)
\left( 9 A(K^0) +9 A(K^+) +9 A(\pi ) +3 A(\eta ) \right)
\nonumber \\
&& +\left( T_1(K^0) + T_1(K^+) \right)
\left( 7 A(K^0) -7 A(K^+)
+2\sqrt{3}\theta_0( A(\eta ) - A(\pi ) )\right)
\nonumber \\
&& +6 T_1(\pi ) \left( A(K^0) - A(K^+) \right)
\Biggr] \nonumber \\
+g_{\mu\nu} &&\left[  {\sqrt{3}\over 24 f^2}  \right]\Biggl[
\left( A(K^0) - A(K^+) \right)
\left(32 A(K^0) +32 A(K^+) +30 A(\pi ) +6 A(\eta ) \right)
\nonumber \\
&&+4\sqrt{3}\theta_0 \left(  A(K^0) + A(K^+) \right)
\left(  A(\eta ) - A(\pi ) \right)
\Biggr]\ ,
\eqnum{3.10}\label{threeten}\end{eqnarray}
(2), from 4(c), (where, owing to the structure of the loop
integrals, only contributions with the central vertex from
the kinetic portion of ${\cal L}^{(2)}$ survive)
\begin{eqnarray}
(q_\mu q_\nu -q^2g_{\mu\nu})
&&\left[  {\sqrt{3}\over 24 f^2} \right]\Biggl[
3q^2\left(  T_1(K^0) - T_1(K^+) \right)
\left(  T_1(\pi ) + T_1(K^0) + T_1(K^+) \right)
\nonumber \\
&& -6\left( T_1(K^0) - T_1(K^+) \right)
\left(  A(K^0) + A(K^+) + A(\pi ) \right)
\nonumber \\
&& -6 \left(  A(K^0) - A(K^+) \right)
\left(  T_1(\pi ) + T_1(K^0) + T_1(K^+) \right)
\Biggr]\nonumber \\
+g_{\mu\nu} &&\left[ - {\sqrt{3}\over 24 f^2} \right]\Biggl[
12\left(  A(K^0) - A(K^+) \right)
\left(  A(\pi ) + A(K^0) +A(K^+) \right)
\Biggr]\ ,\eqnum{3.11}
\label{threeeleven}\end{eqnarray}
(3), from 5(b),
\begin{eqnarray}
g_{\mu\nu}\left[ - {\sqrt{3}\over 24 f^2}  \right]\Biggl[&&
\left( A(K^0) - A(K^+) \right)
\left( 16 A(K^0) +16 A(K^+) +15 A(\pi ) +3 A(\eta ) \right)
\nonumber \\
&& +2\sqrt{3}\theta_0 \left(  A(K^0) + A(K^+) \right)
\left( A(\eta ) - A(\pi ) \right)
\Biggr]\ ,\eqnum{3.12}\label{threetwelve}
\end{eqnarray}
and (4), from Fig.~6,
\begin{equation}
(q_\mu q_\nu -q^2g_{\mu\nu}){4Q\over \sqrt{3}f^2}
\left( m_{K^0}^2-m_{K^+}^2\right)\ .\eqnum{3.13}\label{threenew}
\end{equation}
Adding the results of Eqns.~(3.8) through (3.13), we obtain, for the
full ${\cal O}(q^6)$
contribution to $\Pi^{38}_{\mu\nu}$,
\begin{eqnarray}
\left[\Pi^{38}_{\mu\nu}(q^2) \right]^{(6)} &=&
(q_\mu q_\nu -q^2g_{\mu\nu}) \Biggl[
{i\sqrt{3}\over 2}\lbrace
 (\delta m_{K^+}^2)_{1-loop}
{ \bar{B}(K^+)\over q^2}
- (\delta m_{K^0}^2)_{1-loop}
{ \bar{B}(K^0)\over q^2}\rbrace
\nonumber \\
&& +{4Q\over \sqrt{3} f^2}(m_{K^0}^2 -m_{K^+}^2)
-48i(L_9+L_{10})\left( {\sqrt{3}\over 24 f^2} \right)
\left(  A(K^0) - A(K^+) \right)
\nonumber \\
&&+q^2\left( {\sqrt{3}\over 24 f^2} \right)\lbrace
 T_1(K^0) - T_1(K^+) \rbrace\nonumber \\
&&\times \lbrace 3\left(  T_1(\pi ) + T_1(K^0) + T_1(K^+) \right)
+24iL_9\rbrace\Biggr]\ .
\eqnum{3.14}\label{threethirteen}\end{eqnarray}
As expected, the non-transverse contributions
appearing in (3.8), (3.10), (3.11) and (3.12) have all cancelled.
In Eqn.~(3.14), we may replace the lowest order expressions for
the meson masses appearing in $T_1$, $\bar{B}$ and $A$ with the physical
masses, to the order we are working.  This is not, however,
true of Eqn.~(3.7).  If we wish to combine the
results of (3.7) with those of (3.13), we must re-express
the leading-order squared masses occuring on the RHS of (3.7) as
the differences of the corresponding one-loop expressions
(which can then be set to the physical masses when working
to ${\cal O}(q^6)$ overall) and the ${\cal O}(q^4)$
corrections, $ (\delta m_{K^+}^2)_{1-loop}$,
$ (\delta m_{K^0}^2)_{1-loop}$.
To ${\cal O}(q^6)$
overall it is then sufficient to expand $T_1(m_{K^{+,0}}^2,q^2)$
about the physical values of the squared masses, $M_{K^{+,0}}^2$,
to first order in $ (\delta m_{K^+}^2)_{1-loop}$,
$ (\delta m_{K^0}^2)_{1-loop}$.
The derivative of $T_1(m^2,q^2)$ with respect to $m^2$ which is
required here can be obtained from the expressions in the
Appendix.  The terms first order in
$ (\delta m_{K^+}^2)_{1-loop}$,
$ (\delta m_{K^0}^2)_{1-loop}$
which result turn out to cancel those in (3.13).

Before recording the final ${\cal O}(q^4)$$+$${\cal O}(q^6)$
result for $\Pi^{38}_{\mu\nu}$,
we must discuss the renormalization prescription implicit
in Eqn.~(3.14) (as pointed out above, (3.7) is already
finite and scale-independent).  The loop integrals
$T_1(P)$, $A(P)$ and the LEC's $L_{9,10}$ and $Q$ all
contain divergences as $d\rightarrow 4$.  Those of
$L_{9,10}$ are already known from the renormalization of
${\cal L}_{eff}$ to ${\cal O}(q^4)$\onlinecite{ref1},
and those of $T_1(P)$, $A(P)$ are given in the
Appendix.  Note that, since the vertices arising from ${\cal L}^{(4)}$
and involving $L_{9,10}$ appear in divergent loop
graphs, one must go beyond the expressions for $L_{9,10}$
used in ${\cal O}(q^4)$ calculations
and include the next terms in the Laurent expansions of
these LEC's in terms of the variable
$\bar{\lambda }\equiv {1\over 16\pi^2}\left[
{1\over d-4} -{1\over 2}\left( \log (4\pi)-\gamma_E+1\right)
\right]$,
\begin{equation}
L_k=\mu^{d-4}\left[ L_k^{(1)}(\mu )\bar{\lambda }
+L_k^{(0)}(\mu )+L_k^{(-1)}\bar{\lambda }^{-1}\right]\ ,
\eqnum{3.15}\label{threefourteen}
\end{equation}
where $\mu$ is the dimensional regularization renormalization
scale and $\gamma_E$ is Euler's constant, and, in the more
familiar notation of Ref.\onlinecite{ref1},
\begin{equation}
L_k^{(1)}=\Gamma_k\ \ \ \text{and} \ \ \ L_k^{(0)}=L_k^r\ .
\eqnum{3.16}\label{threefifteen}\end{equation}
One would similarly require, in general, an expression for the
${\cal O}(q^6)$ LEC, $Q$,
of the form
\begin{equation}
Q=\left(\mu^2\right)^{d-4}\left[
Q^{(2)}(\mu )\bar{\lambda }^2 +Q^{(1)}(\mu )\bar{\lambda }
+Q^{(0)}(\mu )\right]\ ,\eqnum{3.17}
\label{threesixteen}\end{equation}
in order to absorb all divergences in two loop calculations.  From (3.14) it
follows that
\begin{equation}
Q^{(2)}(\mu )=0\ \ \ \text{and}\ \ \ Q^{(1)}(\mu )=3\left(
L_9^{(0)}(\mu )+L_{10}^{(0)}(\mu )\right)\ ,
\eqnum{3.18}\label{threeseventeen}\end{equation}
in agreement with the results of Ref.\onlinecite{refGK}, whose
notation we have followed in the expressions above.  Note
that the LEC's $L^{(-1)}_{9,10}$ occur, as claimed earlier,
only at ${\cal O}(q^6)$,
and in the fixed combination,
$\hat{Q}^{(0)}(\mu )=Q^{(0)}(\mu ) -3\left(
L_9^{(0)}(\mu )+L_{10}^{(0)}(\mu )\right)$, with
the ${\cal O}(q^6)$ LEC $Q^{(0)}(\mu )$.  The scale dependence of
the various LEC's is discussed in detail in Ref.~\onlinecite{refGK}
and will not be repeated here.

Given the expressions for the divergent pieces of the loop
integrals $T_1(P)$ and $A(P)$ and the LEC's $L_{9,10}$, one
may now easily verify that the quantities enclosed in braces in
Eqn.~(3.14) contain no divergences as $d\rightarrow 4$.  As
such the ${\cal O}(d-4)$ terms in the expansions of the
$T_1(P)$, $\bar{B}(P)$ and $A(P)$ are not required, and hence have not
been recorded explicitly in the Appendix.  The result is finite
after the renormalization of $Q$ given in (3.17), (3.18) above.  The results of
Ref.~\onlinecite{refGK} for the scale-dependence of the
LEC's also allows one to check that the ${\cal O}(q^6)$ result, (3.14),
is scale-independent.

Using the expressions (3.16), (3.17) and (3.18) (with
$\Gamma_9=-\Gamma_{10}=1/4$ from Ref.~\onlinecite{ref1}),
the explicit form for $T_1(P)$ given in the Appendix, and
rewriting (3.7) in terms of the physical $K^+$ and $K^0$
masses as described above, one obtains the following compact
form for the scalar correlator, $\Pi^{38}(q^2)$, valid to
$6^{\text{th}}$ order in the chiral expansion:
\begin{eqnarray}
\Pi^{38}(q^2) &=& {\sqrt{3}\over 4}(M_{K^0}^2-M_{K^+}^2)_{QCD}
\Biggl[ {-2i\bar{B}(\bar{M}_K^2,q^2)\over q^2}
\Biggl( 1+ \nonumber \\
&&{2q^2\over f^2}\left[ 2L_9^{(0)}
-i\left( \bar{B}_{21}(M_\pi^2,q^2)+2\bar{B}_{21}(\bar{M}_K^2,q^2)
\right) -{1\over 192\pi^2}\log (M_\pi^2\bar{M}_K^4/\mu^6)
\right]\Biggr)\nonumber \\
&&-{(L_9^{(0)}+L_{10}^{(0)})\over 2\pi^2f^2}\left(
1+\log (\bar{M}_K^2/\mu^2)\right)+{16\over 3f^2}\hat{Q}^{(0)}
\Biggr]\ ,\eqnum{3.19}\label{threeeighteen}
\end{eqnarray}
where $\bar{M}_K^2$ is the average of the non-EM portion of
the physical $K^+$ and $K^0$ squared masses,
$(M_{K^0}^2-M_{K^+}^2)_{QCD}$ is the non-EM contribution to the
kaon mass-squared splitting, and the auxillary quantity
$\bar{B}_{21}(M^2,q^2)$ is defined, in terms of $\bar{B}(M^2,q^2)$, by
\begin{equation}
\bar{B}_{21}(M^2,q^2)={1\over 12}\left(1-{4M^2\over q^2}\right)
\bar{B}(M^2,q^2)-{i\over 576\pi^2}\ .\eqnum{3.20}
\label{threenineteen}\end{equation}

For later reference we record here the expression for the imaginary
part of $\Pi^{38}_{\mu\nu}(q^2)$, valid for $q^2<4\bar{M}_K^2$, and to sixth
order in the chiral expansion:
\begin{eqnarray}
\text{Im}\, \Pi^{38}(q^2) = &&
{\sqrt{3} (M_{K^0}^2-M_{K^+}^2)_{QCD}\over 192\pi^2 f^2}
\, \text{Re}\left( i\bar{B}(\bar{M}_K^2,q^2)\right)
\nonumber \\
&&\times\left( 1-{4M_\pi^2\over q^2}\right)^{3/2}\, \theta (q^2-4M_\pi^2)\ .
\eqnum{3.21}\label{threetwenty}\end{eqnarray}
This expression follows straightforwardly from (3.19) and the results
quoted in the Appendix.

\section{Discussion of the Results}\label{sec:4}

The expression (3.19) gives the full result for
$\Pi^{38}_{\mu\nu}(q^2)$,
valid to sixth order in the chiral expansion.  The
functions $i\bar{B}(M^2,q^2)$,
and hence also $i\bar{B}_{21}(M^2,q^2)$,
have cuts beginning at
$q^2=4M^2$.  Since $q^2=4\bar{M}_K^2$ is, presumably, outside
the range of validity of the chiral expansion, the imaginary
part of $\Pi^{38}$ is generated solely by the
$\bar{B}_{21}(M^2_\pi,q^2)$ term in Eqn.~(3.19), in the region
of $q^2$ values of interest to us here.  This contribution
arises from graphs of the form 4(c) having $\pi^+\pi^-$
fields at the $V^3_\mu$ vertex and $K^+K^-$ or $K^0\bar{K}^0$
fields at the $V^8_\nu$ vertex.  Because of the absence of
a lowest order (${\cal L}^{(2)}$) coupling of $V^8_\nu$ to
$\pi^+\pi^-$, the one-loop result for $\text{Im}\Pi^{38}(q^2)$
is zero for $q^2<4\bar{M}_K^2$; the $\pi\pi$ cut first enters
only at two-loop order.

As explained below, all quantities appearing in (3.19) are
previously known, with the exception of the tree-level ${\cal O}(q^6)$
LEC $\hat{Q}^{(0)}$.  Because of significant cancellation between the
$q^2$-independent part of the one-loop contribution and
the term involving $(L_9^{(0)}+L_{10}^{(0)})$, the $\hat{Q}^{(0)}$
term will undoubtedly play a significant role, particularly
near $q^2=0$, and may even be large compared to these other
constant terms individually.  We will comment further on this question
below, but for now will leave $\hat{Q}^{(0)}$
as unknown and investigate the size of the genuine
${\cal O}(q^6)$ loop corrections to the
one-loop result.  In Figure 7 the real and imaginary parts
of $\Pi^{38}(q^2)$ are displayed
(less the constant $\hat{Q}^{(0)}$ contribution
to $\text{Re}\, \Pi^{38}(q^2)$) for $-12\, M_\pi^2\leq q^2\leq
12\, M_\pi^2$, together with the one-loop result (which is purely
real in this range).  One should note that, although the results
are displayed out to $q^2=12\, M_\pi^2$, the corresponding
two-loop expression
for $\Pi^{33}(q^2)$ begins to deviate from experiment
above $q^2\sim (8-9)M_\pi^2$\onlinecite{refGK}.  Moreover, as will
be discussed below,
one should bear in mind that
the range of validity of the two-loop expression for
$\text{Im}\, \Pi^{38}(q^2)$ may not extend as far above threshold as
does the that of the corresponding expression for
$\text{Im}\, \Pi^{33}(q^2)$.

In arriving at the numerical results shown in Fig.~7, we have used the
following input information.  First, we follow standard practice
in taking $f\simeq f_\pi =92.4$ MeV.  Second, having demonstrated
the scale-independence of (3.19), we set $\mu =\bar{M}_K$, which
simplifies the logarithmic terms, and use the values
$L_9^{(0)}(\mu =\bar{M}_K)=0.0073\pm 0.0003$
and $L_{10}^{(0)}(\mu =\bar{M}_K)=-0.0058\pm 0.0003$
(see Refs.\onlinecite{refGK} and \onlinecite{ref18})
(compatible with the resonance saturation result
$L_{10}^{(0)}(\mu =m_\rho )=
-{3\over 4}L_9^{(0)}(\mu =m_\rho )$\onlinecite{ref19,ref20}).  The errors
on $L_{9,10}^{(0)}$ are more significant in the combination
$L_9^{(0)}+L_{10}^{(0)}$,
both due to the large cancellation between the
central values and to the cancellation between the one-loop and
genuine two-loop contributions, resulting in a considerable variation
in the precise value of
$\Pi^{38} (0)$.	 Indeed, including the uncertainties in
$L_9^{(0)}+L_{10}^{(0)}$ by allowing either LEC to vary within
the quoted error bars, one finds the full one-plus-two-loop
curve for the real part of the correlator in Fig.~7 is shifted
up or down by $\sim 0.5\times 10^{-5}$, with little change in
shape (the imaginary part is not affected).  Since, however,
the $\hat{Q}^{(0)}$ contribution is expected to be dominant
(see also the discussion below), this uncertainty is unlikely
to be significant for the correlator as a whole and, as a result,
we have used the central values to obtain the results
shown in the figure.  Finally, we have
obtained the non-EM contribution to the kaon splitting, associated with
$(m_d-m_u)\not= 0$, by subtracting the EM contribution,
where the latter is evaluated
as follows.  In the past the EM subtraction has been made using
Dashen's theorem\onlinecite{ref21}
\begin{equation}
\left( M_{K^+}^2-M_{K^0}^2\right)_{EM}
=\left( M_{\pi^+}^2-M_{\pi^0}^2\right)_{EM}
\simeq \left( M_{\pi^+}^2-M_{\pi^0}^2\right)_{expt}\ ,
\eqnum{4.1}\label{fourone}\end{equation}
a result valid strictly only in the chiral limit.
Recently, arguments have been advanced\onlinecite{ref22,ref23,ref24}
suggesting that the theorem receives significant corrections beyond
leading order and we have, therefore, used the value
\begin{equation}
\left( M_{K^+}^2-M_{K^0}^2\right)_{EM}
\simeq 1.9\, \left( M_{\pi^+}^2-M_{\pi^0}^2\right)_{expt}\
\eqnum{4.2}\label{fourtwo}
\end{equation}
suggested by the analyses of Refs.~\onlinecite{ref23,ref24}
in arriving at $\left( M_{K^+}^2-M_{K^0}^2\right)_{QCD}$.

One significant feature of the results is immediately obvious from
Fig.~7:  despite being higher order in the chiral expansion,
the genuine loop contributions of ${\cal O}(q^6)$
are actually, for most of the $q^2$ range displayed, even larger
in magnitude than the one-loop, ${\cal O}(q^4)$, result.  Moreover, unlike
the ${\cal O}(q^4)$
result, which has a rather small variation with $q^2$, the
${\cal O}(q^6)$ corrections are strongly $q^2$-dependent.  We thus see
that, independent of the value of $\hat{Q}^{(0)}$,
the chiral series for $\Pi^{38}(q^2)$ is poorly converged to
${\cal O}(q^4)$.

The failure of the chiral series for $\Pi^{38}(q^2)$
to be well-converged at one-loop order is actually not a surprise,
given the similarity of the qualitative features of
the ${\cal O}(q^4)$ result, Eqn.~(3.7), to those of the amplitude
for $\eta\rightarrow\pi^0\gamma\gamma$
\onlinecite{ref8,ref9}.  In the latter case, as
for $\Pi^{38}(q^2)$,
the leading, ${\cal O}(q^2)$, contributions are zero,
and the ${\cal O}(q^4)$ LEC's, $\{ L_k^{(0)}\}$, do not
contribute.  Moreover, loop contributions with internal $\pi$
legs are suppressed by a factor of $(m_d-m_u)$, while
loop contributions with internal $K$ legs, which are not so
suppressed, are instead suppressed by the natural smallness of
the $K$ loop integrals.  The latter effect can be easily seen
in the behavior of the loop integral function $\bar{B}(M^2,q^2)$
near $q^2=0$:
\begin{equation}
\bar{B}(M_P^2,q^2)={i\over 96\pi^2}{q^2\over M_P^2}+{i\over 960\pi^2}{q^4\over
M_P^4}+\cdots \ .\eqnum{4.3}\label{fourptthree}\end{equation}
Thus, near $q^2=0$, the $K$ loop integral is, e.g., suppressed by a factor
of $M_\pi^2/\bar{M}_K^2\simeq 0.08$ relative to the corresponding
$\pi$ loop integral.  The result is that the one-loop prediction
for the branching ratio of $\eta\rightarrow\pi^0\gamma\gamma$
is a factor of $\sim 170$ smaller than that
determined experimentally\onlinecite{ref10}.  The reason
for this discrepancy is well-understood, and is relevant to the
case at hand.  As is well-known\onlinecite{ref25,ref19,ref20},
by making standard field choices, one may incorporate heavy
resonances and the pseudoscalar octet into a single effective
chiral Lagrangian.  Integrating out the heavy resonance fields
then produces an effective Lagrangian for the pseudoscalars of
the form given in Section II.  The effect of the heavy resonances is
to produce contributions to the LEC's.  These contributions are
fixed in terms of the parameters describing the couplings of
the pseudoscalars and the heavy resonances in the original,
extended, effective Lagrangian, these parameters, in turn,
being fixed by comparison with experiment.  One then finds that,
where vector and axial vector resonances can contribute to a given
$L_k^{(0)}$, their contributions practically saturate the observed
values\onlinecite{ref19,ref20}.
Thus, for $\eta\rightarrow\pi^0\gamma\gamma$, where the
dominant contribution to the amplitude is known to be due to
vector meson exchange\onlinecite{ref9,refVDM}, the absence of
the $L_k^{(0)}$ in the one-loop result indicates
the complete absence of the dominant contributions to the
amplitude at this order (at least for the interpolating field choice
for the vector mesons implicit in the standard construction).
The effect of the vector mesons, in this case, first appears
in the tree-level constants generated by terms from ${\cal L}^{(6)}$,
and these terms must, therefore, actually dominate the
amplitude\onlinecite{ref9}.  The situation for $\Pi^{38}(q^2)$ is
very similar.  Here we again expect significant, probably
dominant, vector meson exchange contributions, and the absence of
the ${\cal O}(q^4)$ LEC's from Eqn.~(3.7) indicates that these contributions
are not present in the ${\cal O}(q^4)$ result.  It is,
therefore, likely that the $\hat{Q}^{(0)}$
term in (3.19) will be the dominant one, at least at low $q^2$,
especially given the cancellation between the ${\cal O}(q^4)$
result and the genuine loop corrections of ${\cal O}(q^6)$~.  (The
$\hat{Q}^{(0)}$ term, of course, does not contribute to the slope of
$\Pi^{38}(q^2)$ with respect to $q^2$, and this slope will, therefore,
continue to be dominated by the the genuine two-loop
${\cal O}(q^6)$ contributions.)  We will discuss the possibilites for
constraining $\hat{Q}^{(0)}$ in the next section.

We close this section with a brief elaboration of our earlier
comments on the experimental accessibility of the
spectral function, $\rho^{38}(q^2)$, defined by
\begin{equation}
\text{Im}\, \Pi^{38}(q^2)=\pi\, \theta (q^2-4M_\pi^2)\, \rho^{38}(q^2)\ ,
\eqnum{4.4}\label{fourfour}\end{equation}
from a comparison of $e^+e^-\rightarrow \pi^+\pi^-$ and
$\tau^-\rightarrow\nu_\tau\pi^-\pi^0$
data.  The possibility
rests on the presence of ${\cal O}(m_d-m_u)$ isospin-breaking
contributions in the former process, but their absence in the
latter.  As is well-known,
the spectral function of the photon vacuum polarization,
$\rho^{\gamma\gamma}(q^2)$,
which is a linear combination of $\Pi^{33}$, $\Pi^{38}$, and
$\Pi^{88}$, is directly proportional to the measured cross-section
for $e^+e^-\rightarrow hadrons$.  Below $q^2=9M_\pi^2$, only
$\pi^+\pi^-$ intermediate states contribute.  Since the coupling
of $V^8_\mu$ to $\pi^+\pi^-$ is already ${\cal O}(m_d-m_u)$, one
has, to ${\cal O}(m_d-m_u)$,
\begin{equation}
\rho^{\gamma\gamma}(q^2)=\rho^{33}(q^2)+{2\over \sqrt{3}}
\rho^{38}(q^2)\ ,\eqnum{4.5}\label{fourfive}\end{equation}
and the deviation of $\rho^{\gamma\gamma}(q^2)$ from the
value expected based on $\tau^-\rightarrow\nu_\tau\pi^-\pi^0$
and isospin symmetry is, therefore, attributable completely
to $\rho^{38}(q^2)$.  The ratio $r\equiv 2\rho^{38}(q^2)/\sqrt{3}
\rho^{33}(q^2)$ thus represents the accuracy required in
order to be able to extract $\rho^{38}(q^2)$ experimentally.
The expression for $\rho^{38}(q^2)$ which follows from
Eqns.~(3.21) and (4.4) produces values of $r$ of order $2\times 10^{-4}$.
It should be noted that the two-loop result for $\rho^{38}(q^2)$
is likely to be accurate only relatively close to threshold.
This is because the graphs 4(c) which contribute to
$\rho^{38}$ in this range of $q^2$ contain no $\pi\pi$ rescattering.
As is evident from the deviation of the one-loop result for
$\rho^{33}$ from experiment (and hence from the full two-loop result)
even at rather small $q^2$ ($q^2\geq 5M_\pi^2$) -- see Fig.~5 of
Ref.~\onlinecite{refGK}~) -- such effects can be quite significant.
If we use the $\rho^{33}$ results as a guide, such (yet higher order)
effects might enhance $\rho^{38}(q^2)$ by a factor of 2 or so
in the vicinity of $q^2\simeq 8\, M_\pi^2$.  This still leaves
the isospin-breaking correction factor, $r$, at
$\sim 4\times 10^{-4}$, as mentioned in the Introduction, far outside the
reach of current experiments.

\section{The LEC $\hat{Q}^{(0)}$ and the Convergence
of the Chiral Series to Two-Loop Order}\label{sec:5}

The asymptotic behavior of the scalar correlator, $\Pi^{38}(q^2)$
is known from the operator product expansion (OPE) to be,
up to logarithmic corrections \onlinecite{refSVZ}
\begin{equation}
\Pi^{38}(q^2) \simeq{\sqrt{3}\over 8\pi^2}{(m_d^2-m_u^2)\over (-q^2)}\ .
\eqnum{5.1}\label{fiveone}\end{equation}
{}From (5.1), it follows that $\Pi^{38}(q^2)$
satisfies the unsubtracted dispersion relation
\begin{equation}
\Pi^{38}(q^2) =\int_{4M_\pi^2}^\infty ds {\rho^{38}(s)\over s-q^2-i\epsilon}\ .
\eqnum{5.2}\label{fivetwo}\end{equation}
As usual, this means that $\Pi^{38}(q^2)$
and its derivatives with respect to $q^2$ at $q^2=0$ can be
written as negative moments of the spectral function
$\rho^{38}(q^2)$, e.g.,
\begin{eqnarray}
\Pi^{38} (0)&=& \int_{4M_\pi^2}^\infty ds\  {\rho^{38}(s)\over s}
\eqnum{5.3}\label{fivethree} \\
{d\over dq^2}\Pi^{38}(0)&=& \int_{4M_\pi^2}^\infty ds
\ {\rho^{38}(s)\over s^2}
\ .
\eqnum{5.4}\label{fivefour}
\end{eqnarray}
The LHS's of these relations have chiral expansions which involve the quark
masses and the LEC's appearing in ${\cal L}_{eff}$.  If one had
experimental access to the spectral function, these relations (often
called chiral sum rules) would serve to provide information on the
LEC's.  We have not written down the explicit form for these sum
rules since, as pointed out above, $\rho^{38}(q^2)$ is unlikely
to become experimentally available in the near future, but they
are easily constructed from Eqn.~(3.19).  Eqn.~(5.3), in particular,
involves the new, unknown ${\cal O}(q^6)$
LEC $\hat{Q}^{(0)}$~.

Since we cannot realistically hope to constrain $\hat{Q}^{(0)}$
using (5.3), we must look for other ways to estimate its value.
Perhaps the most favorable source of such an estimate would be
the chiral sum rule, analogous to (5.3) above, for the
difference of the $\Pi^{33}$ and $\Pi^{88}$ spectral functions,
as derived in Ref.~\onlinecite{refGK}:
\begin{eqnarray}
\int_{4M_\pi^2}^\infty && \left( {\rho^{33}(s)-\rho^{88}(s)\over s}\right)
\ \ -\ \ {1\over 48\pi^2}{\bar{M}_K^2\over M_\pi^2}\ \ = \nonumber \\
&&-{4M_\pi^2\over 8\pi^2f^2}\log \left( {\bar{M}_K^2\over M_\pi^2}
\right) \left( L_9^{(0)}(\bar{M}_K)+L_{10}^{(0)}(\bar{M}_K)\right)
+ {16\left(\bar{M}_K^2 -M_\pi^2\right)\over 3f^2}\hat{Q}^{(0)}(\bar{M}_K)
\ .\eqnum{5.5}\label{fivefive}\end{eqnarray}
This involves the ${\cal O}(q^4)$
LEC's, $L_9^{(0)}$ and $L_{10}^{(0)}$, which are already rather
well-known, and the at-present-unknown ${\cal O}(q^6)$ LEC
$\hat{Q}^{(0)}$.  An analsyis of the
sum rule (5.5) (in addition to other sum rules involving
$\rho^{33}$ and $\rho^{88}$) is being performed by
Golowich and Kambor, and should provide a useful estimate of
$\hat{Q}^{(0)}$, but at present this analysis has
not been completed.  One might be tempted to follow the path
of estimating $\hat{Q}^{(0)}$
using the resonance saturation hypothesis, which has proven very
successful for the ${\cal O}(q^4)$ LEC's, and
has also been employed in treating the ${\cal O}(q^6)$
LEC's appearing in $\gamma\gamma\rightarrow\pi^0\pi^0$\onlinecite{ref10}
and $\eta\rightarrow\pi^0\gamma\gamma$\onlinecite{ref9}.  In
the present case, however, the application of this method is
more complicated than in previous situations since the term of
interest in ${\cal L}^{(6)}$ (involving the LEC $Q$) is generated
only by graphs involving one ${\cal O}(q^2)$ and one
${\cal O}(q^4)$ vector meson vertex from the
original, extended vector-plus-pseudoscalar effective Lagrangian.
In order to fit the constants which determine the ${\cal O}(q^4)$
vertices, one would have to do a detailed analysis of the
vector meson EM decays which included the pseudoscalar
loop corrections.  While such an analysis would be of interest,
given that the observed vector meson decay constant ratios
show definite deviation from $SU(3)_F$ predictions, it is
not available at present.  We will, therefore, content ourselves
with an alternate estimate based on a QCD sum rule analysis of
the correlator in question.

A sum rule analysis of the related correlator, $\Pi^{3\omega}_{\mu\nu}$,
\begin{equation}
\Pi^{3\omega}_{\mu\nu}(q^2)\equiv i\int d^4x \exp (iq\cdot x)
<0\vert T(V^3_\mu (x)V^\omega_\nu (0))\vert 0>\ ,
\eqnum{5.6}\label{fivesix}\end{equation}
where $V^\omega_\nu\equiv \left( \bar{u}\gamma_\nu u
+\bar{d}\gamma_\nu d\right)/6$,
has recently been performed\onlinecite{refKRM},
updating the earlier analyses of Refs.~\onlinecite{refSVZ}
and \onlinecite{refHHMK}.  Defining
\begin{equation}
\Pi^\phi_{\mu\nu}(q^2)\equiv i\int d^4x \exp (iq\cdot x)
<0\vert T(V^3_\mu (x)V^\phi_\nu (0))\vert 0>\ ,
\eqnum{5.7}\label{fiveseven}\end{equation}
with $V^\phi_\nu\equiv \bar{s}\gamma_\nu s$, and the
scalar correlators $\Pi^{3\omega}(q^2)$, $\Pi^{3\phi}(q^2)$ in
analogy with $\Pi^{38}(q^2)$, one then has
\begin{equation}
\Pi^{38}(q^2)=\sqrt{3}\Pi^{3\omega}(q^2)-{1\over\sqrt{3}}
\Pi^{3\phi}(q^2)\ .
\eqnum{5.8}\label{fiveeight}\end{equation}
$\Pi^{3\omega}(q^2)$ was analyzed in Ref.~\onlinecite{refKRM}
by keeping terms of dimension six or less and working to first
order in $\alpha_{EM}$, $\alpha_s$ and $m_q$ in the
vacuum value of the OPE of the
product of currents, $V^3_\mu (x)V^\omega_\nu (0)$.
Including contributions to $\text{Im}\, \Pi^{3\omega}$ associated
with the $\rho$, $\omega$, $\phi$, $\rho^\prime$ and
$\omega^\prime$ mesons, one finds a very stable analysis for
the correlator and one that, via the unsubtracted dispersion
relation satisfied by $\Pi^{3\omega}$,
can be turned into
a representation of the behavior of the correlator in the
vicinity of $q^2=0$.  Uncertainties in the values of the input
four-quark condensates limit the numerical accuracy of this
representation, but the values of $\Pi^{3\omega}(0)$
and ${d\over dq^2}\Pi^{3\omega}(0)$ appear fixed, certainly
to within a factor of 2\onlinecite{refKRM}.  If a similar
analysis can be performed for $\Pi^{3\phi}(q^2)$, then we may use
(5.8) to provide constraints on our two-loop
representation of $\Pi^{38}(q^2)$.

The analysis of $\Pi^{3\phi}(q^2)$ closely follows that of
$\Pi^{3\omega}$, so we will be rather brief here (the
reader is referred  Refs.~\onlinecite{refSVZ,refHHMK,refSVZprime}
for technical details).  It is immediately
obvious that, owing to the flavor mismatch between the two
currents, the dimension 2 and 4 contributions to the correlator
are absent, at least up to and including terms of ${\cal O}(\alpha_s^2)$.
The leading contributions are then of dimension 6 and,
to ${\cal O}(\alpha_{EM},\alpha_s)$, have the following
form\onlinecite{refSVZ,refHHMK}
\begin{eqnarray}
&&-{\pi\alpha_s\over Q^6}< 0\vert \left( \bar{u}\gamma_\alpha
\gamma_5\lambda^a u
- \bar{d}\gamma_\alpha
\gamma_5\lambda^a d\right)
\left(\bar{s}\gamma^\alpha\gamma_5\lambda^a s\right)\vert 0>
\nonumber \\
&&-{4\pi\alpha_{EM}\over Q^6}< 0\vert \left( -{2\over 9} \bar{u}\gamma_\alpha
\gamma_5 u -{1\over 9}\bar{d}\gamma_\alpha\gamma_5 d\right)
\left(\bar{s}\gamma^\alpha\gamma_5 s\right)\vert 0>\ ,
\eqnum{5.9}\label{fivenine}\end{eqnarray}
where $Q^2\equiv -q^2$.  The mixed flavor condensates appearing in
(5.9) vanish in the standard vacuum saturation approximation and,
being Zweig rule suppressed, are expected to be significantly smaller
than analagous flavor diagonal four-quark condensates in any case.
Estimates for such mixed flavor condensates were made in
Ref.~\onlinecite{refSVZprime} where, for example, it was found that,
for $q=d,s$,
\begin{equation}
{< 0\vert \left( \bar{u}\gamma_\alpha
\gamma_5\lambda^a u\right)\left( \bar{q}\gamma^\alpha\gamma_5\lambda^a q
\right)\vert 0>
\over < 0\vert \left( \bar{u}\gamma_\alpha
\gamma_5\lambda^a u\right)\left( \bar{u}\gamma^\alpha\gamma_5\lambda^a u
\right)\vert 0>}\simeq 0.06\ .
\eqnum{5.10}\label{fiveten}\end{equation}
As such, we should be able to safely neglect (5.9), which approximation
then leads to
\begin{equation}
\Pi^{38}(q^2)\simeq \sqrt{3}\Pi^{3\omega}(q^2)\ .
\eqnum{5.11}\label{fiveeleven}\end{equation}
The results of Ref.~\onlinecite{refKRM} then imply
\begin{eqnarray}
&&\Pi^{38}(0)=(1.5\pm 0.4)\times 10^{-4}\eqnum{5.12} \\
&&{d\over dq^2}\Pi^{38}(0)=(6.2\pm 2.2)\times 10^{-4}\ \text{GeV}^{-2}
\eqnum{5.13}\label{fivethirteen}\end{eqnarray}
where the quoted errors reflect uncertainties in the input values of the
four-quark condensates.  The overall scale of the results in
(5.12) and (5.13) is set by the quark mass difference contribution
to the kaon splitting, which has been determined from the experimental
splitting using the modified version of Dashen's theorem (Eqn.~(4.2)
above).  Using the value determined, instead, by the unmodified
form of Dashen's theorem would lower both numbers by $\sim 20\%$.  It
should be stressed, for the sake of the discussion below,
that attempting to raise the magnitude of the four-quark
condensates sufficiently to lower the values in (5.12) and
(5.13) beyond the lower bounds quoted there
leads to instabilities in the analysis, i.e., the absence
of a ``stability window'' for the extracted resonance parameters
as a function of the Borel mass\onlinecite{refKRM};
such values are, therefore,
inconsistent, at least in the context of the analysis as
presently performed.

If we accept the approximations above, then (5.12) implies
\begin{equation}
\hat{Q}^{(0)}\simeq (8.6\pm 2.7)\times 10^{-5}\ ,
\eqnum{5.14}\label{fivefourteen}\end{equation}
where the errors relect both the uncertainties in (5.12) and those
in the LEC's $L_9^{(0)}$ and $L_{10}^{(0)}$,
though they are dominated by the former.  As argued on
physical grounds above, the $\hat{Q}^{(0)}$
term indeed almost completely dominates $\Pi^{38}(0)$, being
a factor of $\sim 6$ larger than both the ${\cal O}(q^4)$
and ${\cal O}(q^6)$ genuine loop contributions (which are comparable
in magnitude, but opposite in sign).
This, of course, raises questions about
whether yet higher order contributions are necessarily
negligible.  (Recall, e.g., that for $\eta\rightarrow\pi^0\gamma\gamma$,
the full vector meson dominance contribution to the partial width
was a factor of $1.7$ larger than that associated with only
the ${\cal O}(q^6)$ portions thereof\onlinecite{ref9}.)  If we use also
the information contained in (5.13), then it, in fact, appears
that there is good evidence for believing that higher order
contributions must, indeed, be important.  This statement
follows from the observation that the slope of $\Pi^{38}(q^2)$
with respect to $q^2$ is rather well-determined in (3.19), and is
$\sim 9\times 10^{-5}\ \text{GeV}^{-2}$, a factor of at least
$\sim 4$ smaller than that given in (5.13).  Small corrections
to the dimension 6 contributions on the OPE side of the sum
rule are incapable of altering this conclusion.  The statement,
to this order, is of course also independent of the value
of $\hat{Q}^{(0)}$, which makes only a constant contribution to the
scalar correlator.  Thus, if (5.13) is even reasonably
accurate, it clearly demonstrates that even to two-loop order
the chiral series for the correlator is not yet well-converged.
This is somewhat surprising, given the behavior of the
flavor-diagonal correlators to two-loop order\onlinecite{refGK},
but perhaps not as much so as one might, at first, think.
Indeed, the slope of the correlator receives its largest contribution
from the $L_9^{(0)}$
term in (3.19), which contribution is associated with graphs of the
type 4(a) in which the ${\cal L}^{(4)}$ vertex involves $L_9$.
Since the $L_9$ current vertices do not themselves break
isospin, these graphs, like those which contribute to ${\cal O}(q^4)$,
involve only internal $K$ lines and, as a result, the slope is
suppressed by the smallness of the loop integral factor
$\bar{B}(\bar{M}_K^2,q^2)$.  As we saw for the correlator itself,
in going from one- to two-loop order, when one creates the
possibility of internal $\pi$ lines by going to higher chiral
order, such ``suppressed'' contributions can be significantly
enhanced.  At present it is not clear whether or not this
is actually the case here, but it is clearly worth further
investigation.  It will, in particular, be very interesting
to compare the outcome of the analysis of the sum rule (5.5)
with the estimate (5.14) for $\hat{Q}^{(0)}$.
If the two agree, then one will be justified in having increased
confidence in (5.13), as well as (5.12), and the case for the
slow convergence of the chiral series for the correlator
$\Pi^{38}$ to two-loop order will be considerably strengthened.
If not, it would point to some problem with the
truncations usually made in applying the sum rule method, in
the case of the correlator $\Pi^{38}$.

\section{Summary}\label{sec:6}

In summary we have evaluated the mixed-isospin vector current
correlator $\Pi^{38}_{\mu\nu}(q^2)$ to sixth order in the
chiral expansion.  The result is given in compact form in
Eqn.~(3.19), and involves the previously known ${\cal O}(q^4)$
LEC's $L_9^{(0)}$ and $L_{10}^{(0)}$, and a single combination
of ${\cal O}(q^6)$ LEC's, $\hat{Q}^{(0)}$.  The results shows that (1) the
genuine two-loop contributions to the correlator are, over
much of the $q^2$ range considered, larger than
the leading, ${\cal O}(q^4)$
result and that (2) in contrast to the one-loop result,
the two-loop expression has a very strong $q^2$-dependence.
An analysis of the correlator using QCD sum rules yields an
estimate for the ${\cal O}(q^6)$ LEC, $\hat{Q}^{(0)}$,
but at the same time indicates the likelihood that the
two-loop expression for the correlator is not yet fully converged.
Further work on the value of the LEC $\hat{Q}^{(0)}$
is required in order to clarify this issue.

\acknowledgements

The hospitality of the Department of Physics and Mathematical Physics of
the University of Adelaide and the continuing financial support of the
Natural Sciences and Research Engineering Council of Canada are gratefully
acknowledged.
\newpage

\appendix
\section*{Explicit Expressions for the Various Loop Integrals}

We list here the explicit forms of the various loop integrals which
enter the results of Section III.  A detailed discussion of most of
the quantities listed below is given in Ref.~\onlinecite{refHeathnew}
and in Appendix A of Ref.~\onlinecite{refGK}, to which the reader
is referred for details.

The scalar integral, $A(m^2)$, already defined in the text,
is given by
\begin{equation}
A(m^2)=-i\mu^{d-4}\left[ 2m^2\bar{\lambda} +{m^2\log (m^2/\mu^2)\over
16\pi^2}+{\cal O}(d-4)\right]
\eqnum{A.1}\label{a1}\end{equation}
with $\mu$ the regularization scale and $\bar{\lambda}$ as defined
in the text.

Defining the integrals $B(m^2,q^2)$, $B_\mu (m^2,q^2)$ and
$B_{\mu\nu}(m^2,q^2)$ associated with the graphs of Figs.~1(a) and
4(a-c) by
\begin{equation}
\{ B,\ B_\mu ,\ B_{\mu\nu}\}\equiv \int {d^dk\over (2\pi )^d}
{\{ 1,\ k_\mu ,\ k_\mu k_\nu\}\over [k^2-m^2][(k-q)^2-m^2]}\ ,
\eqnum{A.2}\label{a2}\end{equation}
one finds that the integrals $B_\mu$ and $B_{\mu\nu}$ occur
only in the combinations
\begin{eqnarray}
&&T_{\mu\nu}(m^2,q^2)\equiv 4B_{\mu\nu}(m^2,q^2)-2q_\mu
B_\nu (m^2,q^2) -2q_\nu B_\mu (m^2,q^2)+q_\mu q_\nu B(m^2,q^2)
\eqnum{A.3}\label{a3} \\
&&D_{\mu\nu}(m^2,q^2)\equiv 2B_{\mu\nu}(m^2,q^2)-q_\mu B_\nu(m^2,q^2)\ .
\eqnum{A.4}\label{a4}\end{eqnarray}
Explicit calculation shows that $D_{\mu\nu}={1\over 2}T_{\mu\nu}$ (to all
orders in $(d-4)$).  Defining $\bar{B}(m^2,q^2)$ by
\begin{equation}
B(m^2,q^2)=B(m^2,0)+\bar{B}(m^2,q^2)\eqnum{A.5}\label{a5}
\end{equation} one finds for the
relevant independent integrals
\begin{equation}
B(m^2,0)={A(m^2)\over m^2}-{i\over 16\pi^2}\eqnum{A.6}\label{a6}
\end{equation}
\begin{equation}
\bar{B}(m^2,q^2)=\left\{ \begin{array}{ll}
{i\over 8\pi^2}\left[ 1-{1\over 2}\sqrt{1-{4m^2\over q^2}}\
\log \left( 1+\sqrt{1-4m^2/q^2}\over 1-\sqrt{1-4m^2/q^2}\right)\right]
& \\
\qquad\qquad -{1\over 16\pi}\sqrt{1-4m^2/q^2}\ \theta (q^2-4m^2)
& (q^2>4m^2) \\
{i\over 8\pi^2}\left[ 1-\sqrt{{4m^2\over q^2}-1}\ \text{tan}^{-1}
\left( {1\over \sqrt{{4m^2\over q^2}-1}}\right)\right]
&(0<q^2<4m^2) \\
 {i\over 8\pi^2}\left[ 1-\sqrt{1-4m^2/q^2}\ \text{tanh}^{-1}\left(
{1\over \sqrt{1-4m^2/q^2}}\right)\right]& (q^2<0)
\end{array}\right. \eqnum{A.7}\label{a7}
\end{equation}
and, with $T_{\mu\nu}(m^2,q^2)\equiv T_1(m^2,q^2)(q_\mu q_\nu -q^2g_{\mu\nu})
+T_2(m^2,q^2)g_{\mu\nu}$,
\begin{eqnarray}
&&T_1(m^2,q^2)=\left[ 4\bar{B}_{21}(m^2,q^2)+{A(m^2)\over m^2}
+{\cal O}(d-4)\right]\eqnum{A.8} \\
&&T_2(m^2,q^2)=2A(m^2)\eqnum{A.9}
\label{a89}\end{eqnarray}
with $\bar{B}_{21}(m^2,q^2)$ as defined in the text.

The insertion graphs of Fig.~2 involve the integrals
$C(m^2,q^2)$, $C_\mu (m^2,q^2)$ and $C_{\mu\nu}(m^2,q^2)$
defined by
\begin{equation}
\{ C,\ C_\mu ,\ C_{\mu\nu}\}\equiv
\int {d^dk\over (2\pi )^d}{\{1,\ k_\mu ,\ k_\mu k_\nu\}\over
[(k^2-m^2)^2][(k-q)^2-m^2]}\ .\eqnum{A.10}\label{a10}
\end{equation}
These integrals occur in the calculation only in the combination
\begin{eqnarray}
U_{\mu\nu} &&\equiv 4C_{\mu\nu} -2q_\mu C_\nu -2q_\nu C_\mu +q_\mu q_\nu
C\nonumber \\
&&\equiv U_1(m^2,q^2)(q_\mu q_\nu -q^2g_{\mu\nu})+U_2(m^2,q^2)g_{\mu\nu}\ ,
\eqnum{A.11}\label{a11}\end{eqnarray}
where one may show that
\begin{eqnarray}
&&U_1(m^2,q^2)=-{\bar{B}(m^2,q^2)\over q^2}+{\cal O}(d-4)\eqnum{A.12} \\
&&U_2(m^2,q^2)={A(m^2)\over m^2}-{i\over 16\pi^2}+{\cal O}(d-4)\ .\eqnum{A.13}
\end{eqnarray}

As explained in the text, the ${\cal O}(d-4)$ terms in the expressions
for the various integrals do not enter the final result and hence are
not displayed explicitly.

Finally, in order to recast the one-loop result in terms of the physical
kaon masses, one requires the value of the derivative of
$T_1(m^2,q^2)$ with respect to $m^2$, which is readily obtained
from (A.8), (A.1) and the relation
\begin{equation}
{d\over dm^2} \bar{B}_{21}(m^2,q^2)=-{\bar{B}(m^2,q^2)\over 2q^2}
+{i\over 192\pi^2m^2}\ .\eqnum{A.14}\label{a14}
\end{equation}

\begin{figure} [htb]
\centering{\
\epsfig{angle=0,figure=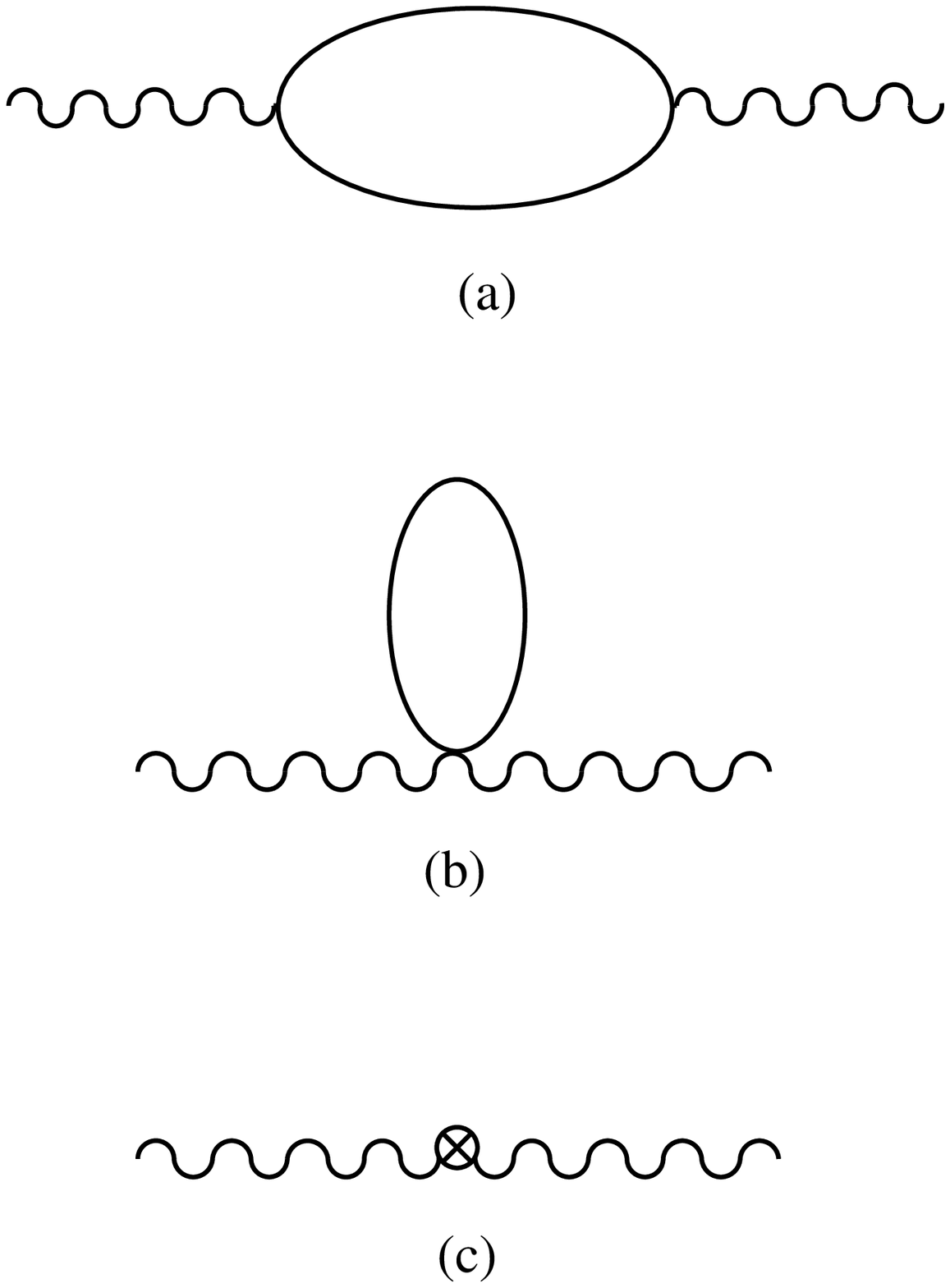,height=9cm}  }
\caption{One-loop contributions to the correlator $\Pi^{38}_{\mu\nu}$}
\label{one}
\end{figure}

\begin{figure} [htb]
\centering{\
\epsfig{angle=0,figure=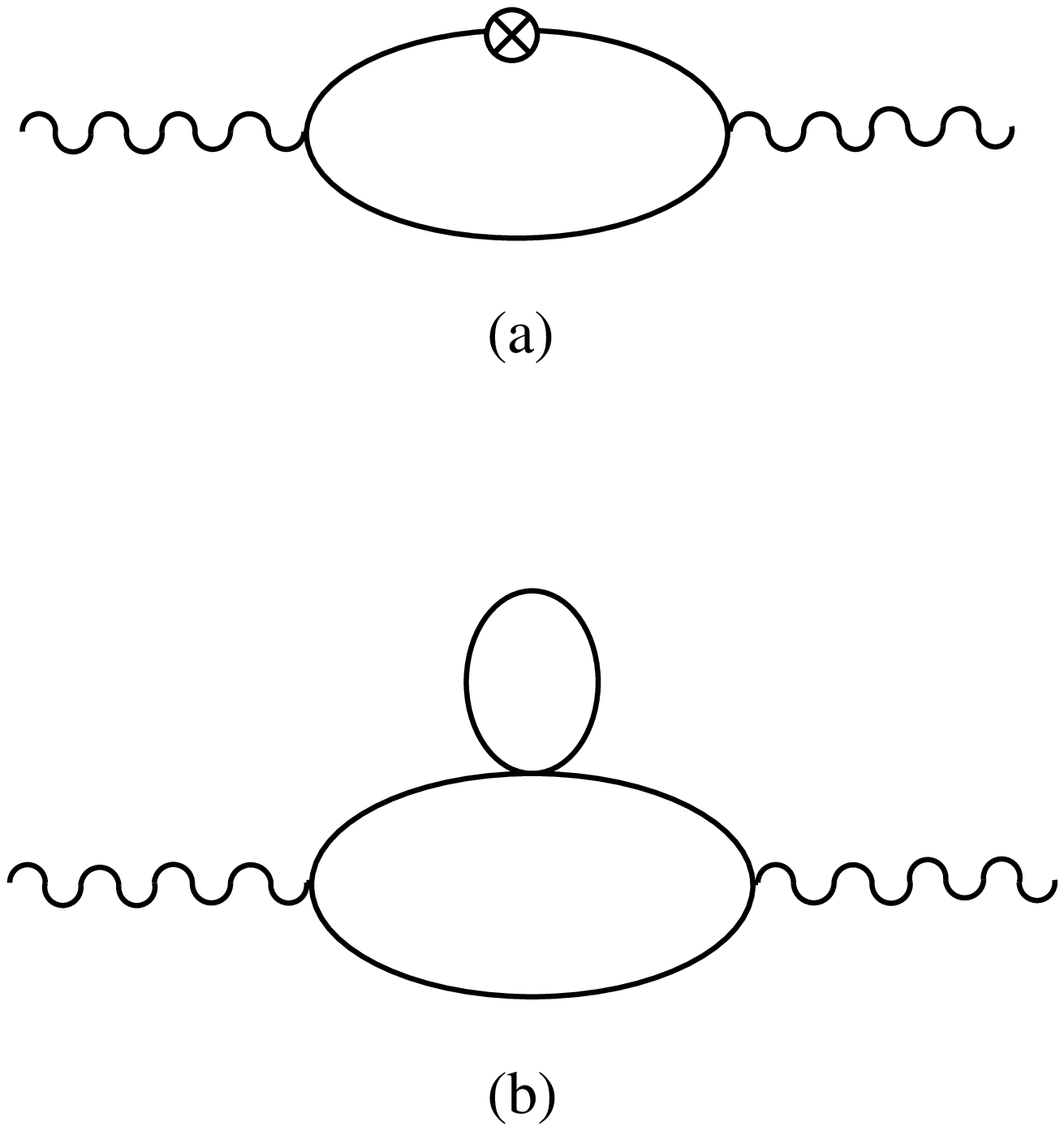,height=9cm}  }
\caption{Non-contact ``insertion'' graphs of ${\cal O}(q^6)$}
\label{two}
\end{figure}

\begin{figure} [htb]
\centering{\
\epsfig{angle=0,figure=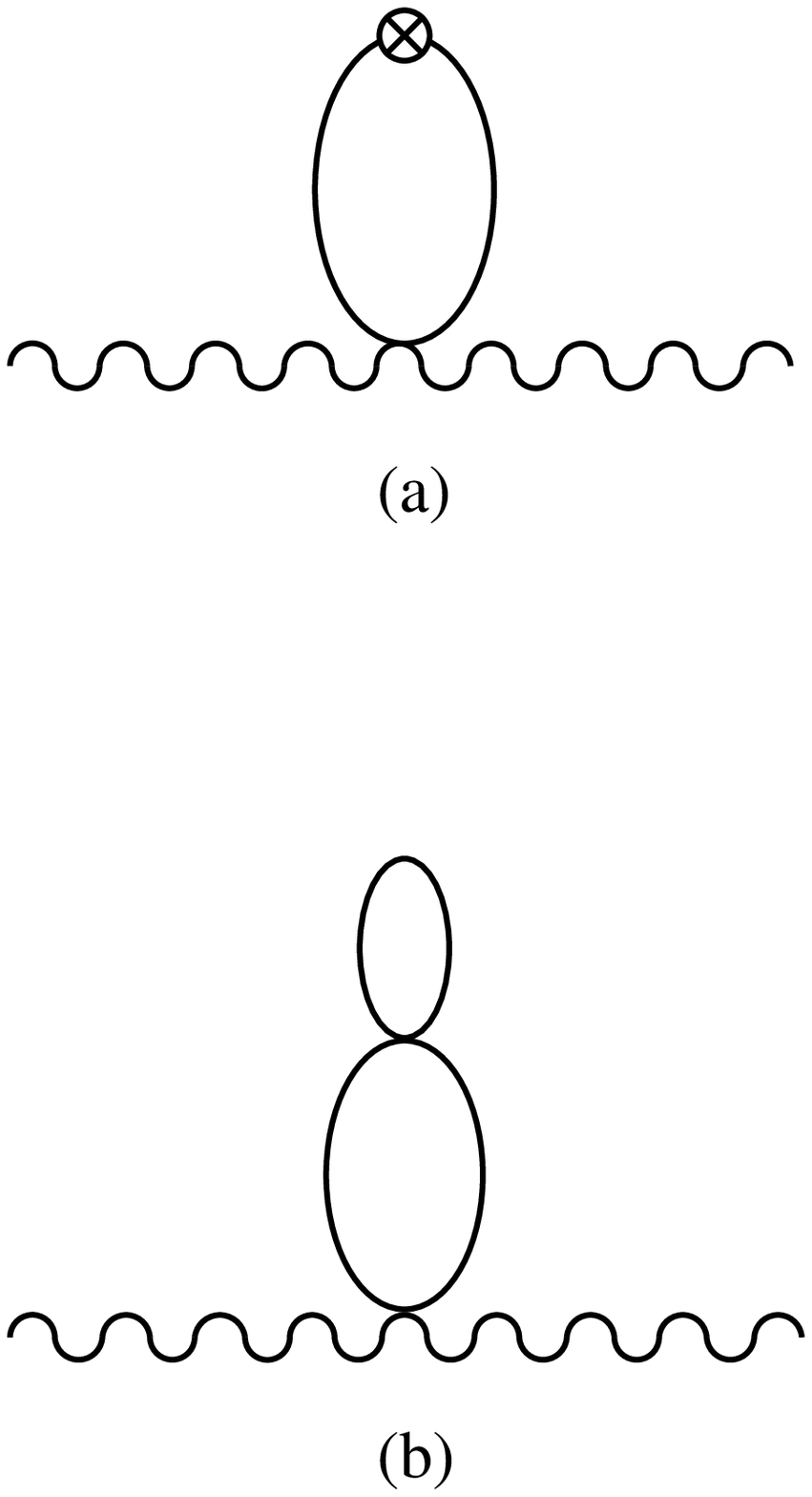,height=9cm}  }
\caption{Contact ``insertion'' graphs of ${\cal O}(q^6)$}
\label{three}
\end{figure}

\begin{figure} [htb]
\centering{\
\epsfig{angle=0,figure=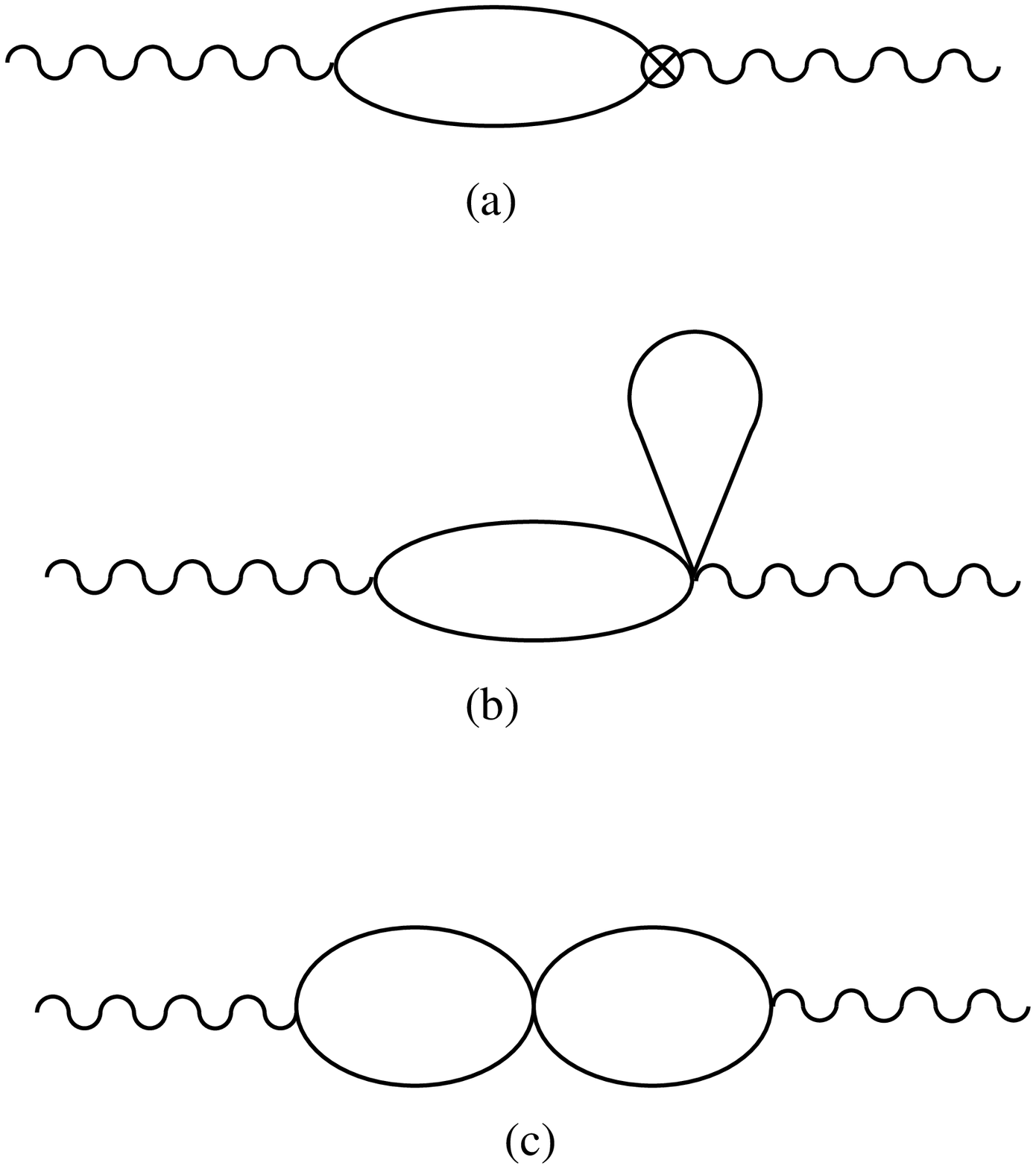,height=9cm}  }
\caption{Non-contact, non-``insertion'' graphs of ${\cal O}(q^6)$}
\label{four}
\end{figure}

\begin{figure} [htb]
\centering{\
\epsfig{angle=0,figure=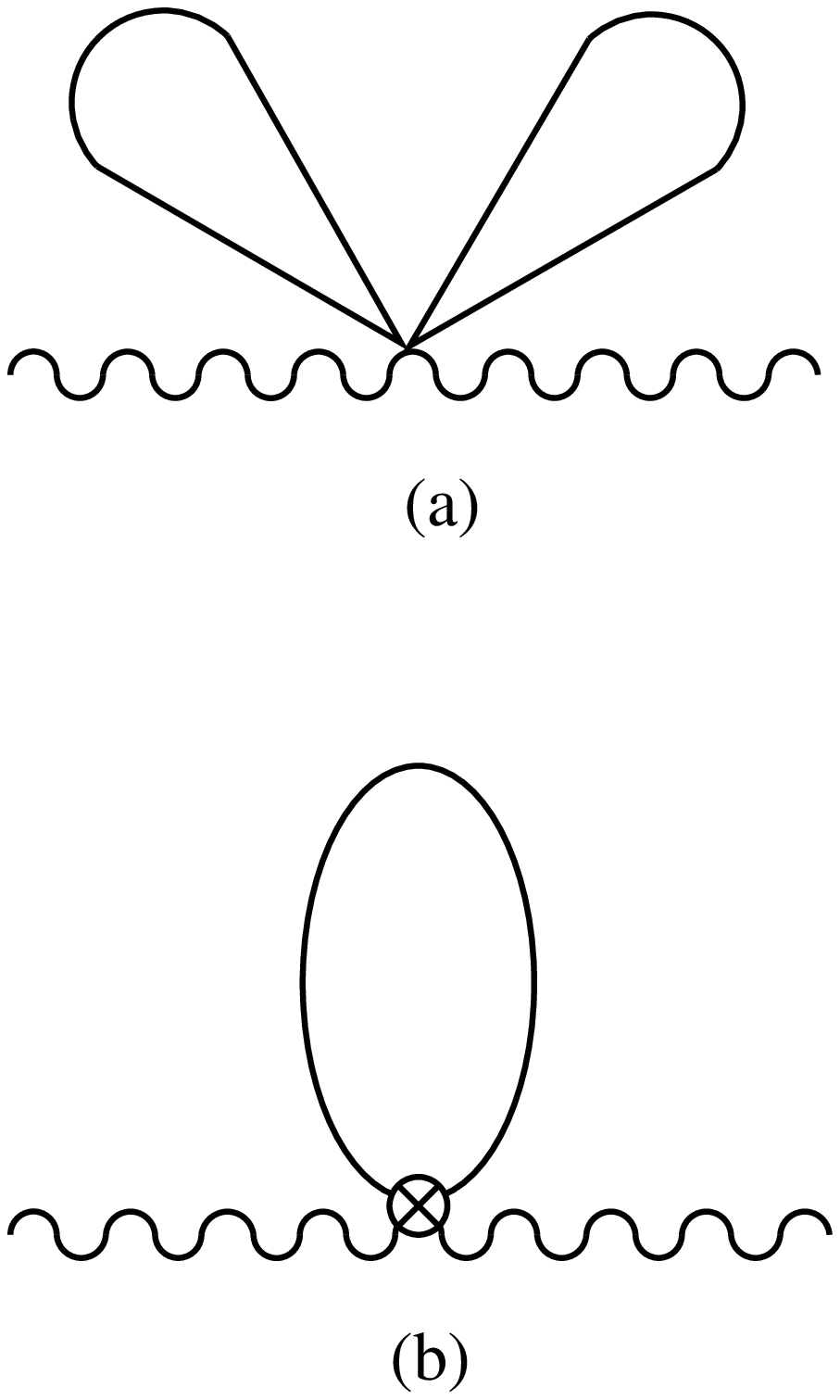,height=9cm}  }
\caption{Contact, non-``insertion'' graphs of ${\cal O}(q^6)$}
\label{five}
\end{figure}

\vskip 2cm

\begin{figure} [htb]
\centering{\
\epsfig{angle=0,figure=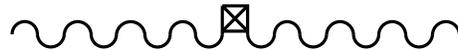,width=6cm}  }
\caption{Tree-level graphs of ${\cal O}(q^6)$}
\label{six}
\end{figure}

\begin{figure} [htb]
\centering{\
\epsfig{angle=0,figure=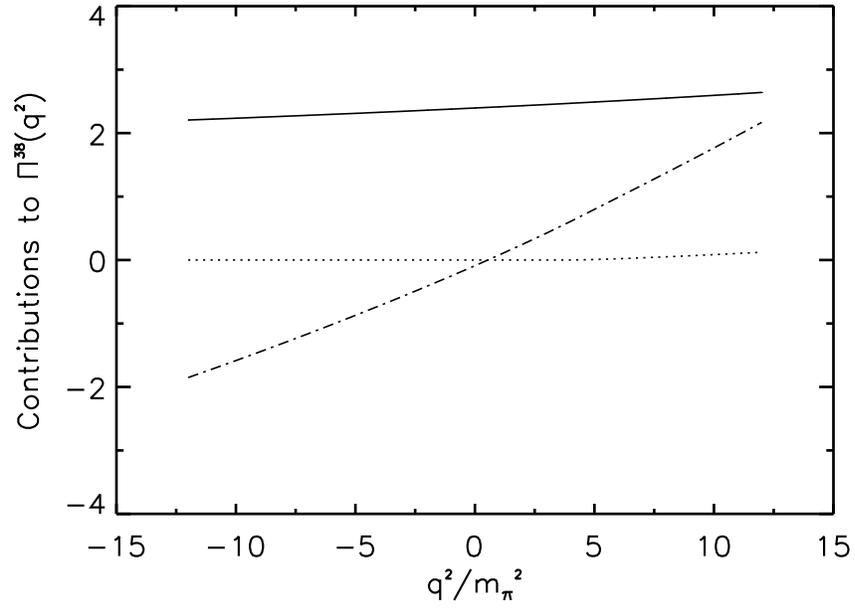,height=9cm}  }
\caption{Loop contributions to the correlator $\Pi^{38}(q^2)$
in units of $10^{-5}$.  The solid line is the one-loop result,
the dotted and dashed-dotted lines the imaginary and real parts
of the full one-loop-plus-two-loop result, respectively.}
\label{seven}
\end{figure}

\end{document}